\documentclass[preprint,aps,amsmath,amssymb]{revtex4}

\usepackage[utf8]{inputenc}
\usepackage{subfig}
\usepackage{amsmath}
\usepackage{textcomp}

\usepackage{graphicx}  % needed for figures

\newcommand{\revision}{}
\DeclareMathOperator{\erf}{erf}

\begin{document}

\title{Autoresonant control of the magnetization switching in single-domain nanoparticles}
\author{Guillaume Klughertz, Paul-Antoine Hervieux, and Giovanni Manfredi}
\email{manfredi@unistra.fr}
\affiliation{Institut de Physique et Chimie des Mat\'{e}riaux de
Strasbourg and Labex NIE, CNRS and Universit\'{e} de
Strasbourg, BP 43, F-67034 Strasbourg, France}

\date{\today}

\begin{abstract}
The ability to control the magnetization switching in nanoscale devices is a crucial step for the development of fast and reliable techniques to store and process information. Here we show that the switching dynamics can be controlled efficiently using a microwave field with slowly varying frequency (autoresonance). This technique allowed us to reduce the applied field by more than $30\%$ compared to competing approaches, with no need to fine-tune the field parameters. For a linear chain of nanoparticles the effect is even more dramatic, as the dipolar interactions tend to cancel out the effect of the temperature. Simultaneous switching of all the magnetic moments can thus be efficiently triggered on a nanosecond timescale.
\end{abstract}

\maketitle

\section{Introduction}
The fast and reliable control of the magnetization dynamics in magnetic materials has been a topical area of research for the last two decades.
In particular, single-domain magnetic nanoparticles have attracted much attention, both for fundamental research on nanoscale magnetism and for potential technological applications to magnetic data storage, which is expected to increase to several petabit/inch$^{2}$ (10$^{15}\rm cm^{-2})$ in the near future \cite{weller,Gubin}. For the fast processing and retrieval of the stored information, a precise control of the magnetization switching dynamics is a necessary requirement \cite{Hillebrands,Back,Gerrits,Schumacher,Seki}.
Single-domain nanoparticles with uniaxial anisotropy possess two stable
orientations of the magnetic moment along the anisotropy axis, separated by an energy barrier proportional to the volume of the particle.
This feature renders them particularly attractive as information-storage units. However, for very small particles the barrier can be of the same order as the temperature, so that the magnetic moment switches randomly between the two orientations under the effect of the thermal fluctuations \cite{weller1}, thus precluding any fine control of the magnetization dynamics.
This phenomenon is known as superparamagnetism.

A potential solution would be to use nanoparticles with high magnetic anisotropy \cite{Sun}.
But an increased anisotropy requires larger fields to reverse the magnetization of the nanoparticle, which is currently difficult to achieve experimentally and causes unwanted
noise. In order to elude this limitation, a microwave field can be combined to the static field \cite{chang, miltat}. For cobalt nanoparticles, it was shown that a monochromatic microwave field can significantly reduce the static switching field \cite{thirion} and that the optimal field should be modulated both in frequency and amplitude using a feedback technique \cite{barros}.
However, the use of such a feedback mechanism can be costly and cumbersome in practical situations.
{\revision Some authors also pointed out that the onset of chaos in the magnetization dynamics can facilitate the reversal of the magnetic moment \cite{daquino}.}

Here, we propose a more effective technique that relies on the concept of autoresonance. This approach was originally devised for a simple nonlinear oscillator (e.g., a pendulum) driven by a chirped force with a slowly varying frequency \cite{fajans01,peinetti,marcus}.
If the driving amplitude exceeds a certain threshold, then the nonlinear frequency of the oscillator stays locked to the excitation frequency, so that the resonant match is never lost (until, of course, some other effects start to kick in).
Importantly, the autoresonant excitation requires no fine-tuned feedback mechanism.

Autoresonant excitation has been observed in a wide variety of environments, including atomic systems \cite{meerson,liu}, plasmas \cite{fajans99a,fajans99b}, fluids \cite{friedland99}, and semiconductor quantum wells \cite{manfredi}. Some authors also noticed the beneficial effect of a chirped pulse on the magnetization dynamics in a nanoparticle \cite{cai,wang,Rivkin}, but lacked the analytical tools provided by the autoresonance theory.
The autoresonance theory was used in the past to study the excitation of high-amplitude magnetization precession in ferromagnetic thin films \cite{Shamsutdinov} and the dynamics of localized magnetic inhomogeneities in a ferromagnet \cite{Kalyakin}. However, those authors did not investigate realistic physical systems and their analysis remained very abstract.

In the present work, we concentrate on a specific physical system that has long been studied experimentally in the past \cite{thirion}, namely single-domain magnetic nanoparticles.
We will show how the autoresonant mechanism can be fully exploited to control the magnetization reversal dynamics in a coherent fashion, on a timescale of a few tens of nanoseconds.
Although this is longer that the picosecond switching time that can be achieved in principle with all-optical techniques \cite{Rasing}, the latter require the use of finely tailored laser pulses and are thus more complex to implement in practice.

{\revision Our analysis takes into account, within the framework of the macrospin approximation, the majority of important
physical mechanisms}, such as the temperature (which is deleterious for the coherent control) and the dipolar interactions between nanoparticles (which turn out to favor coherent switching).
With the proposed method, we are able to reduce the switching field by more than 30\% compared to competing microwave approaches, with no need to fine-tune the field parameters.

\section{Model}
Our treatment can be applied to a variety of physical systems that can be described by a macroscopic magnetization (macrospin).
As a concrete example, we consider an isolated magnetic nanoparticle with uniaxial anisotropy along $\mathbf {e}_z$ in the macrospin approximation
($\lvert \mathbf M \rvert$ is constant), in the presence of an external static field collinear to the anisotropy axis {\revision\footnote{{\revision We also tried other directions of the DC magnetic field. It appears that, as long as $\mathbf B_{DC}$ does not deviate too much from the $z$ axis, the results are unchanged. A systematic analysis of the influence of this angle goes beyond scope of the present work.}}}. An oscillating AC microwave field of varying frequency will constitute the autoresonant excitation. The adopted configuration is sketched on Fig. 1.

\begin{figure}[hbt]
\begin{center}
{\includegraphics[width=.3\textwidth]{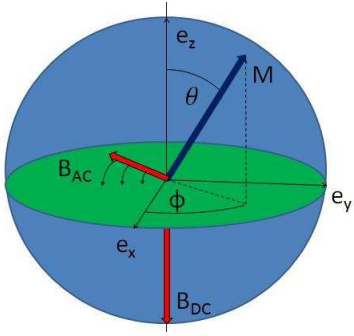}}
  \caption{{\it Color online}. Geometric configuration of the nanoparticle with its magnetic moment ${\mathbf M}(\theta,\phi)$, the static field $\mathbf B_{DC}$, and the time-dependent AC field $\mathbf B_{AC}(t)$. {\revision The case of an AC field rotating in the $(\mathbf e_x, \mathbf e_y)$ plane is shown on the figure}.}
  \label{fig:sub1}
\end{center}
\end{figure}

The evolution of the macroscopic moment $\mathbf M = \mu_S \mathbf m$, of constant amplitude $\mu_S$ and direction {\revision along} $\mathbf m$, is governed by the Landau-Lifshitz-Gilbert (LLG) equation:

\begin{equation}
 {\frac{d \mathbf M}{dt}}=-{\frac{\gamma}{(1+ \lambda ^2)}}(\mathbf M \times  \mathbf B_{eff})\\
 -{\frac{\gamma \lambda}{(1+ \lambda ^2) \mu_S}}[\mathbf M \times (\mathbf M \times  \mathbf B_{eff})], \label{LLG}
\end{equation}
where $\gamma = 1.76\times10^{11} ~\rm(Ts)^{-1}$ is the gyromagnetic ratio, $\lambda=0.01$ the phenomenological damping parameter in the weak damping regime, and $\mathbf B_{eff}$
the effective field acting on the particle. The latter is the sum of the anisotropy field $\mathbf B_{an}=(2KV/\mu_S^2) M_z\mathbf {e}_z$,
the static field $\mathbf B_{DC}= - b_{DC} \mathbf e_z$ and the oscillating microwave field $\mathbf B_{AC}$. Here, $K$ is the anisotropy constant, $V$ is the volume of the nanoparticle, and $\mu_S$ is the magnetization at saturation.
The LLG equation is integrated using the Heun scheme.
We will study the consequences of two kinds of oscillating fields:
a field with fixed direction (along $\mathbf {e_x}$) and varying amplitude
\[
\mathbf B_{AC}^{lin}(t)= b_{AC} \cos[\Omega(t)] \mathbf e_x,
\]
and a field with constant amplitude rotating in the $(\mathbf e_x,\mathbf e_y)$ plane
\[
\mathbf B_{AC}^{rot}(t)= b_{AC} \cos[\Omega(t)] \mathbf e_x + b_{AC} \sin[\Omega(t)] \mathbf e_y,
\]
where $\Omega(t)=2\pi (f_0 t+{\frac{\alpha}{2}}t^2)$, $f_0$ is the initial frequency, and $\alpha$ is the frequency sweeping rate.
Note that the purpose here is not to analyze the different impact of these two types of fields, but rather to show that the autoresonance mechanism is sufficiently general and does not depend on the exact form of the oscillating field.

For the autoresonant excitation to work, the instantaneous frequency $f(t)=f_0 +\alpha t$ must at some instant {\revision become equal to} the linear {\revision resonant frequency of the system $f_r$ \cite{fajans01}, which in our case is given by the precession frequency}. Thus, our strategy is to start from an initial frequency slightly larger than the resonant frequency (i.e. $f_0 > f_r$) and take $\alpha$ negative. When $f(t) \approx f_r$ the magnetic moment starts being captured into autoresonance and its precession amplitude (i.e., the polar angle $\theta$ defined in Fig. 1) keeps increasing, thus entering the nonlinear regime. Thanks to the autoresonant mechanism,
the excitation frequency $f(t)$ remains subsequently locked to the instantaneous nonlinear frequency, which is no longer equal to $f_r$. Therefore, the resonance condition is never lost, and {\revision the precession angle keeps growing until the magnetic moment switches to the $-{\mathbf e}_z$ direction}.

\section{Results for isolated particles}
In order to fix the ideas and analyse the autoresonant excitation in its simplest form, we start with a single isolated nanoparticle, neglecting the effect of temperature and dipolar interactions.
As a typical example \cite{jamet}, we consider a 3nm-diameter Co nanoparticle, with $K=2.2\times10^{5} $J/m$^3$,  $V=14.1~\rm nm^3$, and magnetization at saturation equal to  ${\mu_S}=1500\times1.7\times{\mu_B}=2.36\times10^{-20} $J/T, where $\mu_B$ is Bohr's magneton.
Initially, the magnetic moment $\mathbf M$ is directed along the positive $z$ axis.
Therefore, $f_0$ is determined by computing the {\revision resonant frequency} $f_r=\gamma B_{eff}/2\pi(1+\lambda^2)$ around $\theta=0$ ($\theta$ is the polar angle defined in Fig. 1).
Using $b_{DC}=0.1$T and $ b_{AC}=10$mT, we find $f_r\simeq 4.56$GHz.
As the resonant frequency decreases with growing amplitude, we must choose $\alpha<0$ and $f_0$ slightly above $f_r$. In the following, we shall use $f_0=5$GHz.

\begin{figure}[!ht]
  \centering
  \subfloat[][]{\includegraphics[width=.3\textwidth]{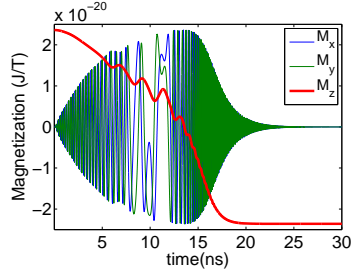}}\\
  \subfloat[][]{\includegraphics[width=.3\textwidth]{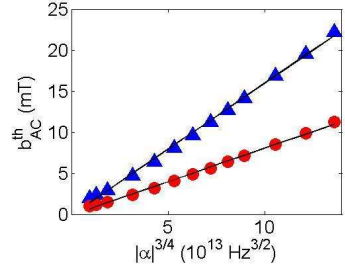}}\\
  \subfloat[][]{\includegraphics[width=.3\textwidth]{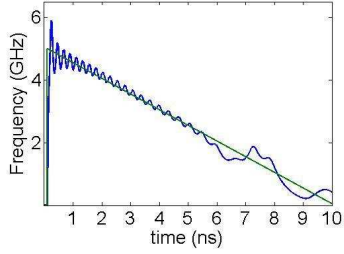}}
  \caption{{\it Color online}. (a) Evolution of the three components of the magnetic moment $\mathbf M$ for a rotating field $B_{AC}^{rot}$. (b) Threshold amplitudes for a rotating field (red circles) and for an oscillating field with fixed direction (blue triangles) as a function of $|\alpha|^{3/4}$. (c) Instantaneous frequencies of the $M_y$ component of the magnetic moment (blue line) and of the applied rotating field $B_{AC}^{rot}$ (straight green line).}
  \label{fig:sub2}
\end{figure}

Figure \ref{fig:sub2}a shows the evolution of each component of the magnetic moment $\mathbf M$ for a rotating field (for a field parallel to the $x$ axis the result is basically identical).
For both cases, $M_x$ and $M_y$ oscillate in quadrature (this is the precession motion around the effective field $\mathbf B_{eff}$) while growing in amplitude, whereas $M_z$ drops from $+\mu_S$ down to $-\mu_S$.
The magnetization switching occurs on a typical timescale of about 20 ns.

According to the theory \cite{fajans01}, the autoresonant mechanism is activated only if the amplitude of the excitation is above a threshold $b_{AC}^{th} \propto |\alpha|^{3/4}$, which is proportional to the frequency chirp rate $\alpha$.
At zero temperature, the transition to the autoresonant regime around the threshold is very sharp and this scaling law is nicely confirmed by the numerical simulations (Fig. \ref{fig:sub2}b).
We note that a microvawe field rotating in the plane perpendicular to the anisotropy axis is slightly more efficient (i.e., it has a lower threshold) than a field oscillating along a given axis.
Figure \ref{fig:sub2}c displays the instantaneous frequency of the microwave excitation (a straight line, since the frequency varies linearly with time) together with the instantaneous frequency of the precessing magnetic moment. Both frequencies stay closely locked together, in accordance with the autoresonant mechanism. {\revision The instantaneous frequency was computed with an algorithm based on the Hilbert transform \cite{goswami}}.

Assuming that the amplitude is larger than $b_{AC}^{th}$, the switching time is determined by the frequency sweeping rate $\alpha$. Once the magnetic moment is captured into
autoresonance, its nonlinear precession frequency is locked to the instantaneous excitation frequency $f(t)=f_0+\alpha t$ (remember that $\alpha<0$).
If we define the switching time $\tau$ as the time it takes for the moment to cross the energy barrier and knowing that the frequency vanishes at
the top of the barrier {\revision {\footnote {\revision At the top of the energy barrier, the precession reverses from counter-clockwise to
clockwise, thus the (instantaneous) precession frequency goes through zero.}}}, we find $\tau=-f_0/ \alpha$. Therefore, if we want the moment to switch rapidly we need a large sweeping rate $\alpha$. However,
increasing the value of $\alpha$ also increases the required microwave field (see Fig. \ref{fig:sub2}b).
Beyond a certain value of $\alpha$, one would lose the benefit of field reduction provided by the autoresonance mechanism.

{\revision
Our switching times can be compared to other methods, such as ballistic magnetization reversal \cite{He,Bazaliy}, which relies on a DC magnetic field that is switched on and off very rapidly. Ballistic reversal can be achieved in sub-nanosecond times, but requires a much larger field ($>1 \rm T$), and the pulse duration must be within a tight time window, although the latter can be broadened using a spin-polarized current when dealing with large magnetic objects \cite{Zhang}.

In contrast, our approach is not dependent on any form of feedback control, nor a precise tailoring of the external magnetic field (static
or oscillating) and, being based on a resonant phenomenon, requires only small magnetic fields.
As mentioned above, the autoresonant reversal time could also be shortened by using a larger chirp rate, at the expense of a stronger applied AC field.
}

\section{Temperature effects}
{\revision So far, we have only considered the zero-temperature (deterministic) case. In this section, we study the influence of thermal effects on the magnetization reversal.
In isolated single-domain magnetic nanoparticles, the magnetization reversal by thermal activation is well described by the N\'{e}el-Brown model \cite{neel,brown}. According to this model, the thermal fluctuations cause the magnetic moment to undergo a Brownian-like motion about the axis of easy magnetization, with a finite probability to flip back and forth from one equilibrium direction to the other.
The N\'{e}el-Brown model is well validated experimentally -- see \cite{wernsdorfer} for the case of 25 nm cobalt nanoparticles, and \cite{respaud} for smaller nanoparticles (1-2 nm).

However, the temperatures that we consider here ($T < 20$ K) are not large enough to produce this flipping effect, so that for all cases that we study the magnetization is initially (almost) aligned with the $z$ axis. Nevertheless, even if they are not capable of reversing the magnetization by themselves, thermal effects still have an influence on the efficiency of the switching technique, as we shall see in the forthcoming paragraphs.
}

For an isolated single-domain particle, Brown  \cite{brown} proposed to include the thermal fluctuations by augmenting the external field with a fluctuating field $\tilde{\mathbf b}(t)$ with zero mean and autocorrelation function given by:
\begin{equation}
\langle \tilde{b}_i (t) \tilde{b}_j (t') \rangle = \frac {2 \lambda k_B T} {(1+ \lambda^2) \gamma \mu_S} \delta_{ij} \delta (t-t'),
\end{equation}
where $i,j$ denote the cartesian components $(x,y,z)$, $\delta_{ij}$ is the Kronecker symbol (meaning that the spatial components of the random field are uncorrelated),
and $\delta (t-t')$ is the Dirac delta function, implying that the autocorrelation time of $\tilde{ \mathbf b}$ is much shorter than the response time of the system. The temperature is thus proportional to the autocorrelation function of the fluctuating field.

At finite temperature, the thermal fluctuations drive the magnetic moment away from the $z$ axis and bring it to a randomly distributed orientation $(\theta_0, \phi_0)$ before the autoresonant field is activated.
The initial amplitudes $\theta_0$ will then be described by a Rayleigh distribution $f(\theta_0)=\frac{\theta_0}{\sigma^2}\exp\left(-\frac{\theta_0^2}{2\sigma^2}\right)$ where $\sigma$ is the scale parameter of the distribution.
This randomness in the initial distribution creates a finite width in the transition to the autoresonant regime, so that the threshold is no longer sharp as in the zero-temperature case. This behavior was already observed in celestial dynamics \cite{wyatt,quillen} and
superconducting Josephson resonators \cite{naaman}.
Note that the thermal fluctuations are active all along the simulations, although their main effect is to randomize the magnetization direction {\em before} the autoresonant field has had time to act. During the autoresonant excitation the thermal effects are present, but their effect is negligible compared to the oscillating field,
{\revision at least for the range of temperatures considered here ($T< 20 \rm K$).}

This effect can be quantified by the capture probability $P(b_{AC})$, defined as the probability for a magnetic moment to switch under the action of an autoresonant field of amplitude $b_{AC}$ (Fig.~\ref{fig:sub3}a). Following the calculations detailed in Appendix~\ref{App:AppendixNoise}, one can write this probability as
\begin{equation}
P(b_{AC})=-\frac {1} {4} \erf\left(\frac {c_0 - b_{AC}} {\sqrt{2} \kappa \sigma}\right)\left[\erf\left(\frac {c_0 - b_{AC}} {\sqrt{2} \kappa \sigma}\right) + 2\right] + \frac {3} {4},\label{cap_proba}
\end{equation}
where $c_0$ is the threshold amplitude for $\theta_0=0$, and $\kappa$ is a numerically determined constant.
The finite-temperature transition is no longer sharp, but instead displays a certain width $\Delta b_{AC}$, which is mathematically defined as the inverse slope of $P(b_{AC})$ computed
at the inflexion point of the curve.
It is also possible to derive an analytical expression for the width $\Delta b_{AC}$ as a function of the temperature and the volume of the nanoparticle (see Appendix~\ref{App:AppendixNoise} for details). One obtains:
\begin{equation}
\Delta b_{AC} \propto \sqrt{\frac{k_B T}{V}}. \label{transit_width}
\end{equation}
{\revision
We note that this dependence is the same as the one obtained from the N\'{e}el-Brown model \cite{neel,brown} for the fluctuating magnetic field arising from the random motion of the magnetic moment under the effect of the temperature. }
\begin{figure}[!ht]
  \centering
  \subfloat[][]{\includegraphics[width=.3\textwidth]{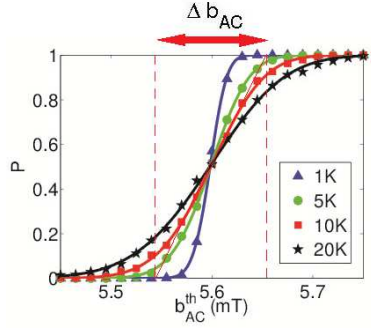}}\\
  \subfloat[][]{\includegraphics[width=.35\textwidth]{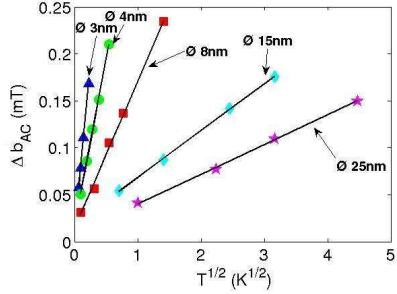}}\\
  \subfloat[][]{\includegraphics[width=.35\textwidth]{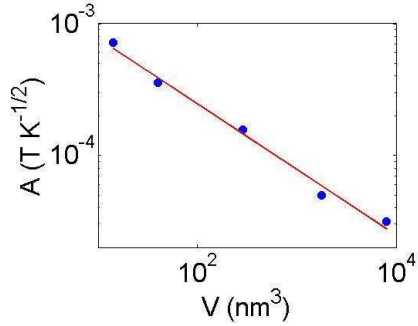}}\\
  \caption{{\it Color online}. (a) Probability to capture the moment into autoresonance as a function of the microwave amplitude for a 25nm-diameter nanoparticle and different temperatures. {\revision Symbols}: numerical simulations. Solid lines: theoretical results from Eq. \eqref{cap_proba}. {\revision The transition width $\Delta b_{AC}$ is shown for the case with $T=10 \rm K$}. (b) Threshold transition width versus $T^{1/2}$ for various diameters. (c) Volume dependence of the parameter $A$ defined as $\Delta b_{AC} = A T^{1/2}$ .
The straight line has a slope equal to $-1/2$.
}
  \label{fig:sub3}
\end{figure}

The capture probability curves of Fig.~\ref{fig:sub3}a are fitted using the analytical expression of Eq. (\ref{cap_proba}) {\revision (the fitting parameter is the product $\kappa \sigma$)}, with excellent agreement between the simulation data and the analytical estimate.
Figure~\ref{fig:sub3}b shows that the transition width scales as the square root of the temperature, as predicted by Eq. (\ref{transit_width}), but the proportionality constant (i.e., the slope) depends on the volume of the nanoparticle.
Plotting the slope as a function of the volume, it can be easily verified that $\Delta b_{AC} \propto V^{-1/2}$, thus confirming both scalings of Eq. (\ref{transit_width}).
Therefore, increasing the size of the nanoparticle diminishes the effect of the temperature on the transition width, making the autoresonant switching observable at experimentally reachable temperatures.

{\revision The above results are of course limited by the applicability of the macrospin approximation, which will cease to be valid for large enough volumes. Nevertheless, for nanoparticles of size 15-30 nm (which covers the range considered in our study), Wernsdorfer and co-workers \cite{wernsdorfer} found that the macrospin approximation is still acceptable.
The validity of the macrospin approximation was also estimated in Ref. \cite{Gubin}; for cobalt nanoparticles, it is expected to break down for a diameter larger than roughly 32 nm (see Table 6.1 in Ref. \cite{Gubin}).
}

\section{Dipolar interactions}
All the preceding results were obtained in the case of a single isolated nanoparticle. For an assembly of densely-packed nanoparticles, dipolar interactions may play a significant role, as was proven in recent numerical simulations \cite{kesserwan}.
The effect of dipole-dipole interactions on the relaxation time and, more generally, on the reversal process has been studied in several works, both theoretical \cite{Hansen,Otero,Dejardin} and experimental \cite{Morup,Dormann,Farrell,Hillion}. Nevertheless, it is still a controversial issue, as opposite dynamical switching behaviors have been reported.
\begin{figure}[!ht]
  \centering
  \includegraphics[width=.35\textwidth]{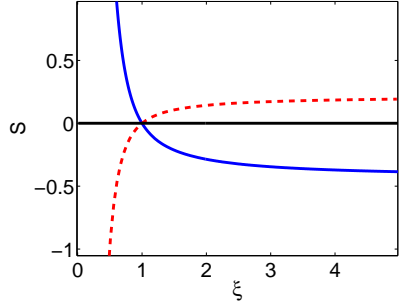}
  \caption{{\it Color online}. The $S(\xi)$ function for an assembly of nanoparticles with the easy axes oriented along the $\mathbf e_z$ direction (solid blue line) or normal to the $\mathbf e_z$ direction (dashed red line).}
  \label{fig:sub4}
\end{figure}

Here, we consider an assembly of interacting particles regularly distributed on a lattice with sites located at $\mathbf r=d_1 (n_1 \mathbf e_x + n_2 \mathbf e_y ) + d_2 (n_3 \mathbf e_z)$,
where $d_1$ and $d_2$ are the centre-to-centre distances between particles in the $(\mathbf e_x,\mathbf e_y)$ plane and in the $\mathbf {e_z}$ direction, respectively,
and $n_1$, $n_2$ and $n_3$ are integers not simultaneously equal to zero.
{\revision
The assembly is supplemented by a number of identical ``replicas" in order to minimize the effect of the boundaries. }

At the instant of capture, the moments are close to the $\mathbf e_z$ axis, and in the case of an
$\mathbf e_z$-oriented assembly of nanoparticles, the dipolar field acting on each moment is also oriented along $\mathbf e_z$. In this configuration, the dipolar interactions can be taken into account via a self-consistent mean dipolar field \cite{denisov,denisov1}
\(
\mathbf B_D= 8 (\mu_0/4\pi) S(\xi) d_1^{-3} \overline{M}_z \mathbf e_z
\)
that acts on all the
nanoparticles. Here, $\overline{M}_z$ is the $z$ component of the mean magnetic moment of the system and $S(\xi)$ is a structure
function describing the geometry of the assembly, defined as:
\begin{equation}
S(\xi) = \frac{1}{8} \sum_{n_1,n_2,n_3} \frac{2 \xi^2 n_3^2 - n_1^2 - n_2^2}{(n_1^2 + n_2^2 + \xi^2 n_3^2)^{5/2}},
\end{equation}
with $\xi \equiv d_2/d_1$.
The sign of $S$ determines if the moments will order ferromagnetically ($S>0$, for essentially 1D systems where $\xi<1$) or antiferromagnetically ($S<0$, for 2D systems where $\xi>1$). The behavior of the function $S(\xi)$ is shown in Fig.~\ref{fig:sub4} (solid blue line).

\begin{figure}[!ht]
  \centering
  \subfloat[][]{\includegraphics[width=.3\textwidth]{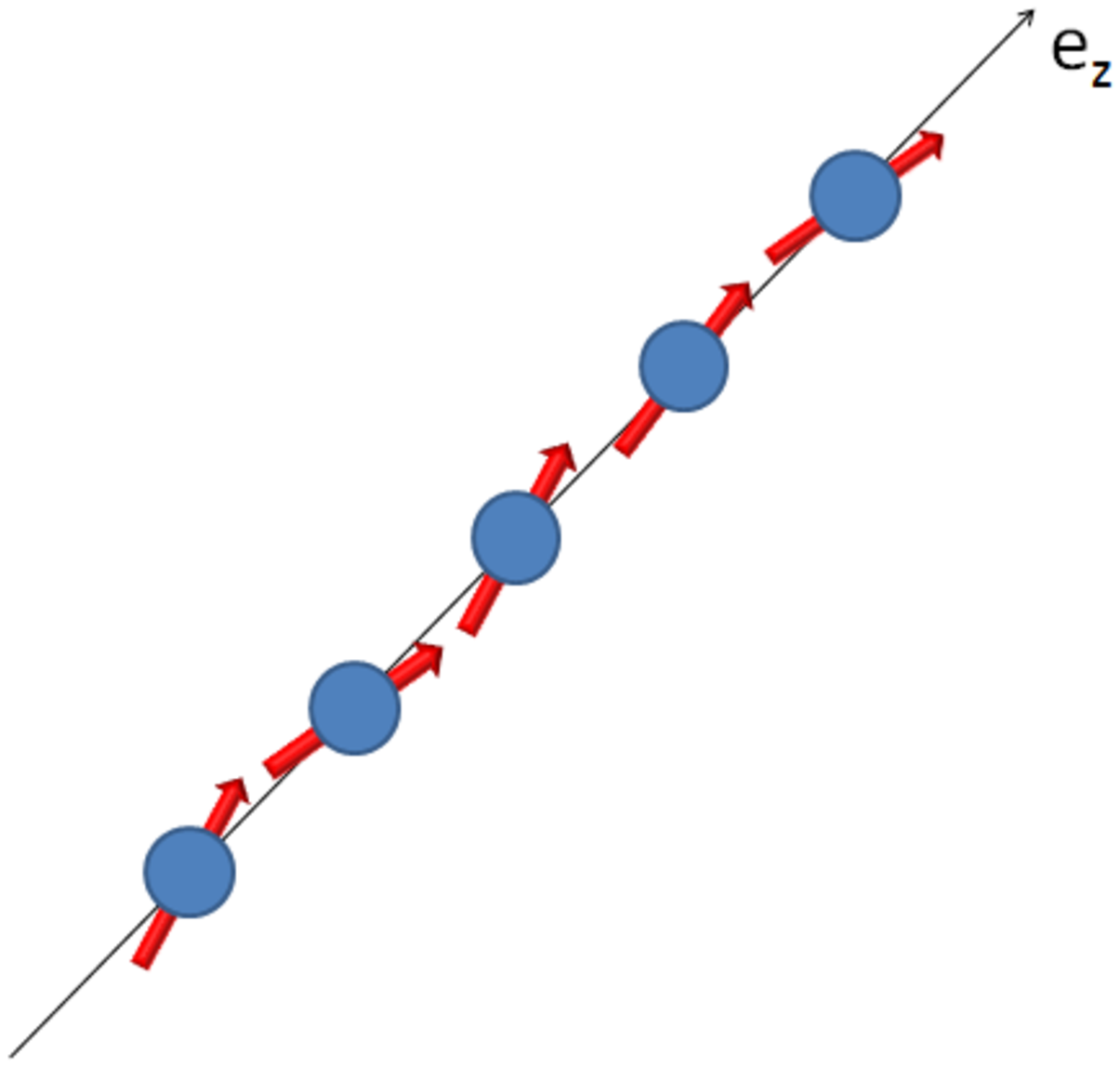}}\\
  \subfloat[][]{\includegraphics[width=.3\textwidth]{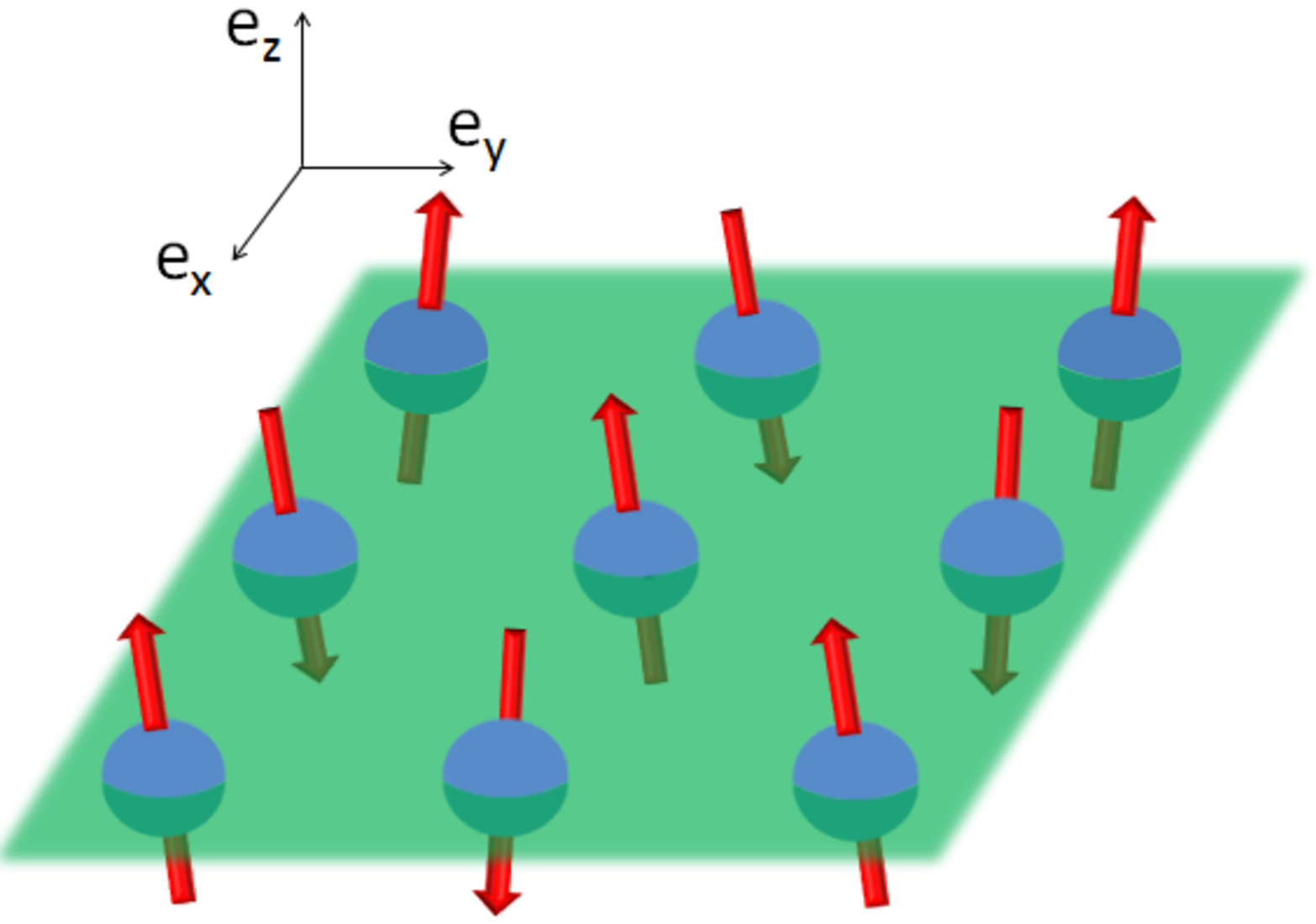}}\\
  \caption{{\it Color online}. Schematic view of the nanoparticle configurations. (a) One-dimensional chain along the $\mathbf e_z$ axis: the particles are ordered ferromagnetically, with small tilts off the $\mathbf e_z$ axis due to the temperature. (b) Two-dimensional film in the $(\mathbf e_x,\mathbf e_y)$ plane with antiferromagnetic order.}
  \label{fig:sub5}
\end{figure}

We studied two typical distributions of the nanoparticles: a 1D chain oriented along the $\mathbf {e_z}$ axis ($\xi\rightarrow 0,S\rightarrow\infty$), and a two dimensional
configuration in the $(\mathbf e_x,\mathbf e_y)$ plane ($\xi\rightarrow\infty$, $S\rightarrow -1.129$). These configurations are represented schematically in Fig. \ref{fig:sub5}.
{\revision
Intermediate values of $\xi$ correspond either to a set of stacked 2D arrays  (when $\xi>1$), or a set of parallel 1D chains of nanoparticles (when $\xi<1$).}

{\revision
It must be noted that the above considerations only apply to the cases where the easy axes are oriented along the $z$ directions, i.e., parallel to the chain in the 1D case, and normal to the plane in the 2D case.
In other cases, the nature of the magnetic equilibrium may be different.
For instance, a chain of particles with their easy axes oriented perpendicularly to the chain direction would behave antiferromagnetically; conversely, a 2D array of nanoparticles with the easy axes parallel to the plane of the array would display a ferromagnetic behavior at equilibrium.
Indeed, for such cases, the function $S(\xi)$ displays an opposite behavior compared to the configurations of Fig. \ref{fig:sub5}, namely it is negative for $\xi<1$ (1D antiferromagnetic) and positive for $\xi>1$ (2D ferromagnetic) (see Fig. 4, red dashed line).

Nevertheless, in our mean-field approach all the information about the geometry is included is the function $S(\xi)$. Different configurations that have the same value of $S$ behave identically in the mean-field limit.
Therefore, in order to fix the ideas, in the remainder of this section we will focus on the geometries sketched in Fig. \ref{fig:sub5}, which described by the solid blue curve on Fig. 4.
}

\subsection{Two-dimensional planar configuration}
\begin{figure}[!ht]
  \centering
  \includegraphics[width=.3\textwidth]{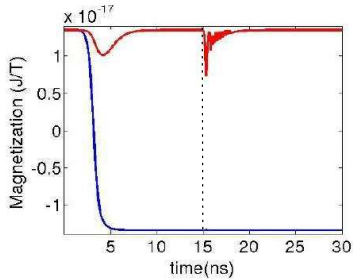}
  \caption{{\it Color online}. Magnetization dynamics in a planar assembly. Evolution of the $M_z$ component for a nanoparticle whose moment has reversed due to the dipolar interactions (blue) and for a nanoparticle whose moment stays aligned along $+ \mathbf {e_z}$ (red curve). The oscillating field is switched on at $t=15$ ns (vertical dashed line).}
  \label{fig:sub6}
\end{figure}

{\revision The autoresonance mechanism is ineffective in a 2D configuration  where the easy axes of the particles are oriented in the direction normal to the plane}.
The reason is that such a planar configuration naturally leads to an antiferromagnetic order, with half the moments pointing in the $+ \mathbf e_z$ direction, and the other half in the $- \mathbf e_z$ direction.
We have preformed a numerical simulation in order to illustrate this fact (see Fig. \ref{fig:sub6}),
{\revision
using an assembly of nanoparticles with diameter equal to 25 nm and interparticle distances $d_1=50$ nm and $d_2\rightarrow\infty$.}

We start, as usual, from a state where all moments are parallel to $+ \mathbf e_z$, and then let the dipolar interactions create the anti-ferromagnetic order. Very quickly ($t \approx 5-10 ~\rm ns$), the dipolar interactions create an antiferromagnetic order: half of the moments reverse, while the other half stays parallel to $+ \mathbf {e_z}$.
We look at two representative moments: one that has switched to the $- \mathbf e_z$ direction (blue curve in Fig.~\ref{fig:sub6}) and one that has not (red curve).

At $t=15$ ns, once the magnetic order is settled, the rotating field is switched on and tries to capture and maintain the moments in autoresonance.
The magnetic moment that had reversed to the $- \mathbf e_z$ direction (blue curve in Fig. \ref{fig:sub6}) is maintained in that direction by the rotating autoresonant field, because this moment naturally precesses in the opposite way, so that the
rotating field tends to counteract its precession.
But the same rotating field is also unable to reverse a moment that points in the $+ \mathbf e_z$ direction (red curve), because the interaction with its four nearest neighbours (all pointing
along $- \mathbf e_z$) destroys the phase-locking even for $b_{AC}$ well above the threshold ($\approx 10b_{AC}^{th}$). This moment can be driven slightly away from its original $+ \mathbf e_z$ axis (see the red curve at $t \approx 15-20$ ns), but soon the dipolar interactions become too strong and restore the antiferromagnetic order.
The autoresonant technique is therefore inefficient for a planar assembly of magnetic nanoparticles.

\subsection{One-dimensional linear chain}
In contrast, a linear chain of nanoparticles {\revision with the easy axes oriented along the chain} displays a ferromagnetic behavior, because $S>0$ for $\xi < 1$ (see Fig.~\ref{fig:sub4}, solid blue line). At equilibrium, all the moments are oriented parallel to the $z$ direction, apart from small fluctuations due to the temperature (Fig.~\ref{fig:sub5}a). Therefore, there is a chance that the autoresonant mechanism may work in this type of configuration.
In order to fix the ideas, we concentrate on a 1D chain of magnetic moments with fixed interparticle distance in the $(x,y)$ plane ($d_1=1\mu$m) and vary the distance $d_2$  along the $z$ axis from $1\mu$m to $26$nm, so that $\xi$ varies between 0.026 and 1.

\begin{figure}[h]
\centering
 \includegraphics[scale=0.4]{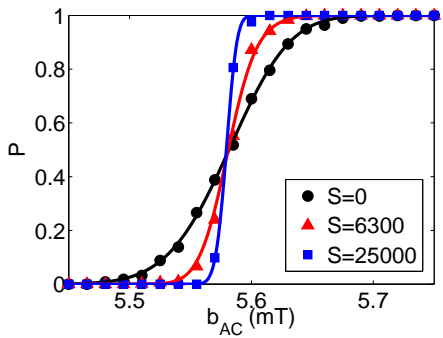}
 \caption{{\it Color online}. Capture probability as a function of the microwave amplitude for a chain of 25nm-diameter nanoparticles at $T=10$K, for different interparticle distances: $d_2=1\mu\rm m$ (black circles), $36\rm nm$ (red triangles), and $26\rm nm$ (blue squares). The corresponding values of $S(\xi)$ are indicated on the figure.}
	    \label{fig:sub7}
\end{figure}

The effect of the dipolar interactions on the autoresonant switching is summarized in Fig.~\ref{fig:sub7}, which shows the capture probability as a function of the microwave amplitude for a 25nm-diameter nanoparticle at $T=10$K, for different interparticle distances along the $z$ axis.
With decreasing interparticle distance (i.e., increasing dipolar interactions), the transition width shrinks, as was also observed for other physical systems \cite{barth}. The dipolar interactions can almost completely erase
the effect of the temperature for dense enough particle assemblies, as in the case with $d_2=26$nm in Fig.~\ref{fig:sub7}.

In reality, the self-consistent dipolar field does not stay aligned along $z$ during the reversal, so that the mean-field approximation fails at some point. However, its main effect occurs {\em before} the magnetic moment has reached the top of the barrier, and until then the approximation is valid. In other words, the dipolar interactions help the moments to be captured into autoresonance; once they are captured, the mean-field approximation is no longer accurate, but then the effect of the external field far outweighs that of the dipolar field, so that the error is irrelevant. Exact calculations for $N$ interacting moments (much more computationally demanding) also confirmed the above picture.

The dipolar interactions also slightly lengthen the switching time by increasing the effective potential barrier, which makes the resonant frequency $f_r$ higher. As we have to choose $f_0>f_r$, the switching time $\tau \sim f_0/ \alpha$ also increases,
but still remains of the order of 10-100 ns for all the cases studied.

\section{Conclusions}
We investigated the possibility to reverse the magnetization of a single-domain Co nanoparticle by combining a static field
with a chirped microwave field. Using the LLG equation, we produced convincing evidence in favor of the autoresonance mechanism and showed that a chirped microwave field with a very small amplitude (a hundred times smaller than the static field) can efficiently reverse the magnetization.

Previous attempts \cite{barros} to use a microwave field to reverse the magnetization showed that the microwave excitation should be modulated both in frequency and amplitude.
Using the same parameters and configuration as in  \cite{barros}, but exploiting the autoresonance mechanism,
we were able to reverse the magnetic moment with $b_{DC}=0.1$T and $b_{AC}=11$mT, reducing the amplitudes of both fields by roughly $30\%$.
For an assembly of many nanoparticles, dipolar interactions can have a significant impact on the switching dynamics. The most favorable configuration is that of a linear chain of nanoparticles, for which the dipolar interactions can drastically reduce the effect of the temperature .

Compared to competing microwave techniques that use sophisticated feedback mechanisms, the autoresonance approach requires no fine tuning of the excitation parameters and thus appears to be a promising candidate for the fast control of the magnetization dynamics in densely-packed assemblies of magnetic nanoparticles.

\bigskip

{\bf  \noindent Acknoledgments}\\
We thank Dr. Jean-Yves Bigot for several helpful suggestions. We acknowledge the financial support of the French ``Agence Nationale de la
Recherche" through the project Equipex UNION, grant ANR-10-EQPX-52.

\appendix

\section{Autoresonance transition with thermal noise}\label{App:AppendixNoise}

As discussed in more details in the article, the presence of noise broadens the transition to the autoresonant regime. In the main text, we mentioned that the transition width $\Delta b_{AC}$, Eq. \ref{transit_width}, is proportional to $\sqrt{T/V}$, where $T$ is the temperature and $V$ is the volume of the nanoparticle.
Here, we derive the full expression for $\Delta b_{AC}$.

The critical amplitude $b_{AC}^{th}$, beyond which the phase-locking is complete, is periodic in $\phi_0$ (the azimuthal angle at the onset of the oscillating field) and can therefore be expanded
in a Fourier series \cite{barth}:
\begin{equation}
b_{AC}^{th}=c_0 + \kappa \theta_0 \cos (\phi_0 + \delta) + ...
\end{equation}
where the angles $(\theta_0, \phi_0)$ define the initial moment orientation,
$c_0$ is the threshold amplitude for $\theta_0=0$, and $\kappa$ can be determined numerically.
For small initial amplitudes, one can restrict the expansion to the lowest order in $\theta_0$.
The capture probability (i.e., the probability to activate and maintain the autoresonant mechanism until magnetization reversal) can then be defined as
\begin{equation}
P(b_{AC})=\int_0^\infty P(\theta_0, b_{AC})f(\theta_0) d\theta_0,
\end{equation}
where $P(\theta_0, b_{AC})=(1/ \pi) \arccos [(c_0 -b_{AC})/(\kappa \theta_0)]$ and
$f(\theta_0)=\frac{\theta_0}{\sigma^2}\exp\left(-\frac{\theta_0^2}{2\sigma^2}\right)$ is the Rayleigh distribution characterizing the initial amplitudes
resulting from the thermal noise. Actually it is more convenient to calculate
\begin{equation}
\frac {\partial P(b_{AC})} {\partial b_{AC}}=\int_0^\infty \frac {\partial} {\partial b_{AC}}(P(\theta_0, b_{AC})f(\theta_0)) d\theta_0,
\end{equation}
This calculation yields:
\begin{equation}
\frac {\partial P(b_{AC})} {\partial b_{AC}}=\frac {1} {\sqrt{2\pi} \kappa \sigma} e^{\frac {-(c_0 - b_{AC})^2} {\sqrt{2} (\kappa \sigma)^2}}[1 - \erf\left(\frac {-(c_0 - b_{AC})} {\sqrt{2} \kappa \sigma}\right)],
\end{equation}
Then, taking the antiderivative:
\begin{equation}
 P(b_{AC})=-\frac {1} {4} \erf\left(\frac {c_0 - b_{AC}} {\sqrt{2} \kappa \sigma}\right)\left[\erf\left(\frac {c_0 - b_{AC}} {\sqrt{2} \kappa \sigma}\right) + 2\right] + C,
\end{equation}
Now, knowing that $\lim\limits_{b_{AC} \to \infty} P(b_{AC}) = 1$ we find the value of the integration constant $C=\frac {3} {4}$. Finally :
\begin{equation}
P(b_{AC})=-\frac {1} {4} \erf\left(\frac {c_0 - b_{AC}} {\sqrt{2} \kappa \sigma}\right)\left[\erf\left(\frac {c_0 - b_{AC}} {\sqrt{2} \kappa \sigma}\right) + 2\right] + \frac {3} {4}.
\end{equation}
The derivative of $P(b_{AC})$ gives a slope $P'=(\sqrt {2 \pi} \sigma \kappa)^{-1}$ at $b_{AC}=c_0$, whose inverse is defined as the
transition width $\Delta b_{AC}$.

One can derive an analytical expression of the mean square displacement of the moment during a short time $\Delta t$ under the influence of the temperature,
which is widely used in Monte Carlo simulations \cite{nowak,chubykalo,lyberatos}. One can write the linearized LLG equation for the normalized moment ${\bf m}$ in the form:
\begin{eqnarray}
        \frac{dm_x}{dt} &=& L_{xx} m_x + L_{xy} m_y,\\
        \frac{dm_y}{dt} &=& L_{yx} m_x + L_{yy} m_y.
\end{eqnarray}
with
\begin{eqnarray}
        L_{xx}=L_{yy}&=&-{\frac{\gamma \lambda}{(1+ \lambda ^2)}} m_z B_{z, eff},\\
        L_{xy}=L_{yx}&=&-{\frac{\gamma }{(1+ \lambda ^2)}}B_{z, eff}.
\end{eqnarray}

Also, close to the local energy minimum $E_0$, one can write the energy $E = E_0 + \Delta E$ where $\Delta E = \frac {1}{2} \sum_{i,j} C_{ij} m_i m_j$ is the energy increase due to
the small fluctuations of $m_x$ and $m_y$. Because of the interactions between the different subsystems the energy matrix $C_{ij}$ is nondiagonal, but it is possible to
perform a transformation to the normal coordinates of the system and write $C$ as a diagonal matrix $\tilde C$. One can then write:
\begin{equation}
\Delta E = \frac {1}{2} (\tilde C_{xx} m_x^2 + \tilde C_{yy} m_y^2)
\end{equation}
with $\tilde C_{xx}=\tilde C_{yy}= \frac{\mu_S}{m_z}B_{z, eff}$.
The correlation matrix of the random forces $\mu_{ij}$ can be defined from $\tilde C_{ij}$ and $L_{ij}$ as $\mu_{ij}=-k_B T \sum_k{(L_{ik} \tilde C_{kj}^{-1} + L_{jk} \tilde  C_{ki}^{-1})}$.
Supposing that $m_z^2\approx1$, the calculation yields:
\begin{equation}
\mu_{xx} = \mu_{yy} = \frac {2 k_B T \lambda \gamma} {(1+ \lambda^2) \mu_S}.
\end{equation}
Finally, one finds the mean square displacement by integrating over a finite time interval $\Delta t$:
\begin{equation}
\langle m_{x}^2 \rangle =\langle m_{y}^2 \rangle = \mu_{xx} \Delta t = \frac {2 k_B T \lambda \gamma} {(1+ \lambda^2) \mu_S} \Delta t.
\end{equation}

On the other hand, the expectation value of $\theta_0$ computed from the distribution $f(\theta_0)$ is $\langle \theta_0 \rangle= \sigma\sqrt{\pi/2}$.
As $\langle \theta_0 \rangle = \arcsin \left(\sqrt {\langle m_{x,0}^2 \rangle + \langle m_{y,0}^2 \rangle}\right)$, one can write the transition width as a function of
the different system parameters:
\begin{equation}
\Delta b_{AC} = \frac {1} {P'} = 4 \kappa(\Delta t) \sqrt{ \frac {k_B T \lambda \gamma} {(1+ \lambda^2) M_S V} ~\Delta t},
\end{equation}
where we have used the expansion $\arcsin(x) = x + o(x^2)$, valid for small initial amplitudes, and $\mu_S=M_S V$.

Note that $\Delta b_{AC}$ depends on $\Delta t$ only in the transient regime.
Once the initial amplitude distribution has reached the Rayleigh equilibrium, the numerically determined ``constant" $\kappa(\Delta t)$ exactly balances the term $\sqrt{\Delta t}$, so that $\Delta b_{AC}$ does not depend on $\Delta t$ anymore.

\end{document}